\documentstyle[prl,aps,floats,twocolumn,graphics,epsfig]{revtex}
\author{F. Krzakala$^1$ and O.C. Martin$^{1,2}$}

\def\(({\left(}
\def\)){\right)}
\def\[[{\left[}
\def\]]{\right]}

\title{Absence of an equilibrium ferromagnetic spin glass phase in three dimensions}
\address{
 $^1$ Laboratoire de Physique Th\'eorique et Mod\`eles Statistiques,
b\^at. 100, Universit\'e Paris-Sud, F--91405 Orsay, France. \\
 $^2$ Service de Physique de l'\'Etat Condens\'e,
Orme des Merisiers --- CEA Saclay, 91191 Gif sur Yvette Cedex, France.}
\date{\today}
\draft

\begin{document}
\twocolumn[ \hsize\textwidth\columnwidth\hsize\csname
@twocolumnfalse\endcsname 
\maketitle

\begin{abstract}
Using ground state computations, we study the transition from a spin glass 
to a ferromagnet in 3-d spin glasses when changing the mean
value of the spin-spin interaction. We find good evidence
for replica symmetry breaking up till the 
critical value where ferromagnetic ordering sets in, and
no ferromagnetic spin glass phase. This phase
diagram is in conflict with
the droplet/scaling {\it and} mean field theories of spin glasses.
We also find that the exponents of 
the second order ferromagnetic transition 
do not depend on the microscopic Hamiltonian,
suggesting universality of this transition.
\end{abstract}
]

The nature of the frozen phase in finite dimensional 
spin glasses~\cite{Young98} is still not understood. Many
studies seem to support the many valley picture predicted by the 
mean field approach~\cite{MezardParisi87b,MarinariParisi99b} in 
which replica symmetry 
is broken (RSB). Nevertheless other
approaches~\cite{FisherHuse86,BrayMoore86}
are not excluded; one open question is 
whether spin glass ordering can 
co-exist with ferromagnetism. The different
theoretical predictions here are in conflict.
Moreover, this question is of experimental relevance: 
it may be possible to test for the co-existence
of the two types of orderings in doped ferromagnets that show
glassy behavior.

In this paper we study the transition from a spin glass to a ferromagnet at
zero-temperature for Ising spin models having nearest 
and next-nearest neighbor interactions in $d=3$. First, 
we find that RSB, which 
is associated with system-size excitations having $O(1)$ energies, 
persists in the presence of an excess of ferromagnetic 
over anti-ferromagnetic bonds. Second, our order parameter for 
RSB vanishes at the same concentration as 
where ferromagnetism sets in; in fact it seems 
that ferromagnetic and spin glass orderings, with or without 
RSB, do not co-exist. These properties are in conflict with
the mean field {\it and} droplet/scaling 
theories~\cite{FisherHuse86,BrayMoore86},
the main theoretical frameworks for spin glasses. Finally, the 
critical exponents and Binder cumulant at criticality
for the ferromagnetic transition are the same for our two models, 
and fully compatible with those found in a similar 
system~\cite{Hartmann99c}, 
giving evidence for universality in this transition.

\begin{figure}
\centerline{\hbox{\epsfig{figure=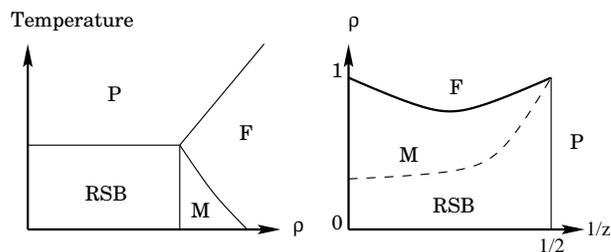,width=8cm}}}
\caption{Left: Phase diagram in a mean-field model ($\rho$ is the excess
concentration of ferromagnetic bonds, c.f. eq.~\ref{eq_H_EA}).
A mixed (M) phase, with both ferromagnetic (F) ordering and replica symmetry 
breaking (RSB), is present. At larger temperature
the system is paramagnetic (P). 
Right: The role of the coordination on the presence of such a mixed
phase in the mean field picture at temperature $T=0$ (artist's view).}
\label{fig_phase}
\end{figure}

\paragraph*{Theoretical expectations ---} 
In the mean field picture, based on the
infinite range Sherrington-Kirkpatrick (SK) 
model~\cite{SherringtonKirkpatrick75},
one has the following sequence at low temperature 
when the concentration of ferromagnetic interactions
is increased (see Fig.~\ref{fig_phase}): (a) an RSB 
phase with no magnetization ($m=0$); (b) a ``mixed''
phase with both RSB 
and $m>0$; (c) a standard ferromagnetic phase. 
The same phase diagram arises in the Random Energy Model~\cite{Derrida81}.
In these models the mixed phase follows from 
the existence of a spin glass phase in a magnetic field; indeed, a non zero 
magnetization creates an effective magnetic field, and thus
mixed phases seem unavoidable in mean field.
One can also consider finite connectivity
generalizations~\cite{VianaBray85,DominicisGoldschmidt89} 
of the SK model where a spin is connected 
at random to a finite number of other spins.
A study of the associated phase diagram  
has been performed~\cite{KwonThouless88} for connectivity $3$ and
again shows the presence of a mixed phase, while the limit
of large connectivities brings us back to the SK model. Now
the combination of spin glass
ordering and very low connectivity most likely inhibits
ferromagnetism, so
the size of the mixed phase should grow as the
connectivity increases.
This is what we illustrate in Fig.~\ref{fig_phase}. If mean field is a 
good guide for three dimensions, then a mixed phase should be 
easier to observe on lattices with large connectivities; for this
reason we shall consider lattices with next-nearest neighbor interactions.

Within the droplet or scaling theories, the 
phase diagram is very different because 
any magnetic field destroys the spin glass ordering. For instance,
within the droplet picture~\cite{FisherHuse86}, the mixed phase
is incompatible with compact droplets. Similarly, within the
scaling picture~\cite{BrayMoore86}, 
having a mixed phase would require a line of neutral fixed points
connecting two unstable fixed points under the
renormalization group, i.e., highly singular and non generic flows.
Not surprizingly, the absence of a mixed phase is explicitly
confirmed in the Migdal-Kadanoff bond-moving 
approach~\cite{SouthernYoung77,MiglioriniBerker98}. This 
picture works very well in low dimensions, and in particular
it agrees with the consensus of no mixed phase in 
dimensions $d \le 2$~\cite{KawashimaRieger97,NishimoriBook01}.
Our focus in this work is the case $d=3$ where the question
of a mixed phase remains open. Note that high temperature
series cannot address this problem as the mixed phase
resides beyond the first boundary of singularities. A good
way to tackle it is via
Monte Carlo; unfortunately, such 
studies have systematically avoided the question of the mixed phase,
focusing instead on ordering from the paramagnetic side.
The only attempt to search for an equilibrium mixed phase we
are aware of is that of Hartmann~\cite{Hartmann99c} who worked
at zero-temperature but was not able
to estimate precisely the boundary of the spin glass phase and
thus made no claims about the existence or not of a mixed phase.

\paragraph*{The models  ---} We consider Edwards-Anderson-like
Hamiltonians
on a $L \times L \times L$ cubic lattice:
\begin{equation}
\label{eq_H_EA}
H_J(\{S_i\}) = (1-\rho) \sum_{<ij>} J_{ij} S_i S_j - \rho \sum_{<ij>} S_i S_j
\end{equation}
where $S_i = \pm 1$ are Ising spins and $\rho$ 
controls the amount of frustration 
in the system. The sums are over all nearest 
neighbor spins in the First Neighbor
Model~(FNM), and over nearest and next-nearest neighbor spins 
(a total of $18$) in the Second Neighbor Model~(SNM). The quenched 
couplings $J_{ij}$ are independent random variables, taken from a Gaussian 
distribution of zero mean and unit variance and we impose periodic 
boundary conditions. This Hamiltonian has 
both a spin glass term and a ferromagnetic term; 
for $\rho=0$ we recover the standard spin 
glass model while for $\rho=1$ we have the usual Ising ferromagnet. 
We use sizes up to $L=12$ for the FNM and up to $L=8$ for the SNM.
At these sizes, the heuristic algorithm~\cite{HoudayerMartin01} 
we use gives the ground state with very high probability so that
our systematic errors are much smaller than our statistical ones.

\paragraph*{Ferromagnetic ordering ---}  
\begin{figure}
\centerline{\hbox{\epsfig{figure=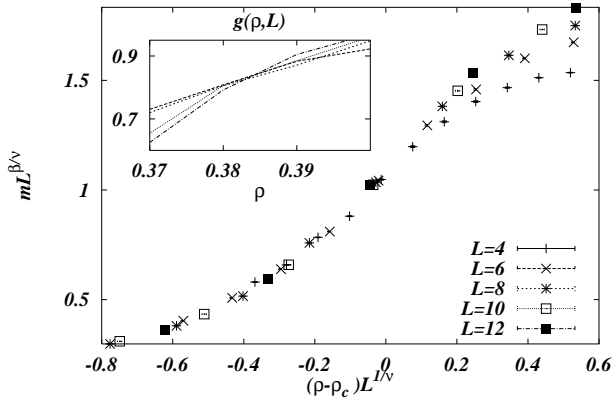,width=8cm}}}
\caption{Scaling plot of $m$ as a function 
of $\rho$ in the First Neighbor Model; ${\rho}_F=0.385$, $\nu=0.9$ and 
$\beta=0.3$. 
The inset shows the Binder cumulant of $m_J$.}
\label{fig_m2_FNM}
\end{figure}
We first focus on the transition from a state where
the magnetization per spin $m$
is zero to a ferromagnetic state ($m\ne 0$). Note that
the Hamiltonian has the global symmetry 
$\{ S_i \} \to \{ - S_i \}$, so $m$ is an order parameter
for the transition. For each instance, we 
determine the magnetization
density $\pm m_J$ (since the $J_{ij}$ are independent continuous
random variables, there are only two ground states, related by symmetry).
Our order parameter is then
\begin{equation}
m = \left[  {\overline{m_J^2}} ~\right]^{1/2}
\end{equation}
where ${\overline{~ \cdot ~} }$ denotes the average over the
disorder variables $\{ J_{ij} \}$. 

For a large excess of
ferromagnetic interactions ($\rho \approx 1$),
$m$ will be close to $1$, while for a small
excess ($\rho \approx 0$), $m$ will go to zero in the
large volume limit.
\begin{figure}
\centerline{\hbox{\epsfig{figure=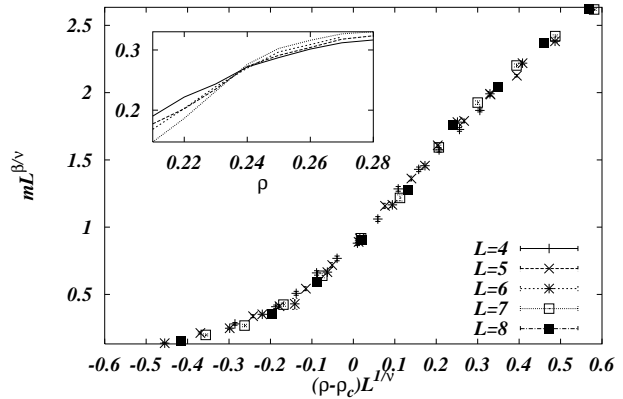,width=8cm}}}
\caption{Scaling plot of the mean value of $m$ 
as a function of $\rho$ in the Second Neighbor Model;
${\rho}_F=0.236$, $\nu=0.9$ and $\beta=0.3$. 
The inset shows the Binder cumulant of $m_J$.}
\label{fig_m2_SNM}
\end{figure}
For both the FNM and SNM, the transition 
appears to be of second order; indeed
our distributions of $|m_J|$ 
always have a single peak whereas a first order 
transition should lead to two peaks at the transition point,
one at $m>0$ and one at $m=0$. (These two peaks
would be generated by those samples that are magnetized and 
those that are not.)
To determine the critical value $\rho_F$,
we have computed the Binder cumulant associated with the
distribution of $m_J$: 
\begin{equation}
g \(( \rho,L \)) =
{\frac{1}{2}} \(( 3- {\frac{\overline {m_J^4}}{{\overline {m_J^2}}^2}} \)) .
\end{equation}
The different $L$ curves cross very well for both models, giving
${\rho}^{FNM}_F=0.385(5)$ and ${\rho}^{SNM}_F=0.236(5)$.
Moreover, applying finite size scaling, we should have
$
g \(( \rho,L \)) =
{\tilde g} \(( L^{1/\nu} \((\rho - \rho_F \)) \))  .
$
Doing so for the order parameter, 
we expect
$ m \(( \rho,L \)) =
L^{-\beta/ \nu} F \(( L^{1/\nu} \((\rho - \rho_F \)) \))$.
The corresponding data collapse is very satisfactory
as can be seen in Fig.~\ref{fig_m2_FNM} and \ref{fig_m2_SNM}.
Furthermore, the values of the critical exponents 
are close for the two models: we find $\beta=0.3(1)$ and $\nu=0.9(2)$.
These values are within the error bars of those
found by Hartmann who studied the case 
where $J_{ij}=\pm 1$~\cite{Hartmann99c}, obtaining
$\beta=0.2(1)$ and $\nu=1.1(3)$. Last but not least, the values
of the Binder cumulant {\it at} the critical point are all
close to one-another.
These results give evidence 
in favor of universality in spin glasses, at least for this 
particular transition.

\paragraph*{Replica symmetry breaking and sponges  ---} 
RSB literally means that $P(q)$, the probability distribution
of spin overlaps taken between equilibrium configurations
is not concentrated as in a ferromagnet. Unfortunately,
estimations of $P(q)$ are plagued by finite size effets. 
Hartmann~\cite{Hartmann99c}
measured the width of $P(q)$ among
ground states of the $\pm J$ model but, because of
subtleties intrinsic to that model, he was not able
to determine the location of the spin glass phase boundary
and thus made no claims about the existence of a
mixed phase. In our work we use a different probe of
RSB that seems to suffer far less from finite size effects.
The starting point is to remark that in the presence of RSB
several macroscopically different valleys
dominate the system's partition function, having 
$O(1)$ differences in free-energy. Extrapolating
such behavior to zero-temperature, we expect
to have system-size excitations above the ground state
costing only $O(1)$ in energy. In our models,
we characterize the
system-size excitations by their topology, following the
procedures developped in~\cite{KrzakalaMartin00}. First, we produce
an excitation by taking at random $2$ spins in the ground state, forcing 
them to change their relative orientation, and recomputing 
the new ground state with that constraint. By definition, 
the spins that are flipped in an excitation form
a connected cluster; if it and its complement ``wind'' around in all 
three $(x,y,z)$ directions, we call it a sponge.
(For instance a cluster winds in the $x$ direction if there exists
a closed walk on it having non-zero winding number in $x$.)
We use the proportion of such sponge-like (topologically non-trivial) events 
as an order parameter for RSB.
In the droplet model, this quantity decays as $L^{-\theta}$ 
so asymptotically there are no large scale $O(1)$ excitations.
As in~\cite{KrzakalaMartin00}, we increased the signal to noise
ratio by first ranking the spins according to their local
field; then we took the two spins at random from
the list's top $25 \%$ (or $50 \%$, both led to the same conclusions).

\begin{figure}
\centerline{\hbox{\epsfig{figure=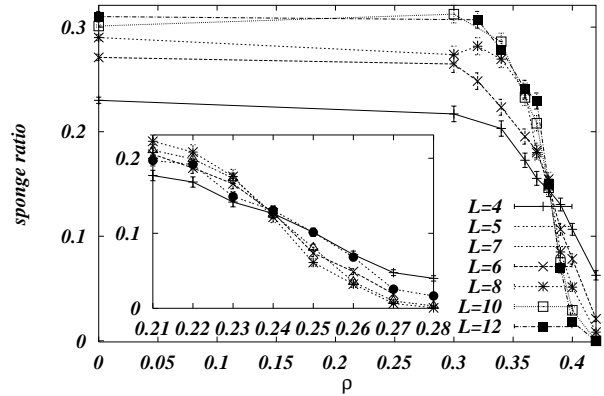,width=8cm}}}
\caption{Fraction of sponge-like excitations versus $\rho$ in the FNM. 
(The inset shows the data for the SNM.) ${\rho}^{FNM}_{{RSB}}=0.380(5)$ (${\rho}^{SNM}_{{RSB}}=0.235(5)$).}
\label{fig_rho_ex}
\end{figure}

We have measured this order parameter
as a function of $\rho$ and $L$.
Our results are very similar in the FNM and the SNM.
Looking at the plot in Fig.~\ref{fig_rho_ex},
there seems to be a phase transition in the FNM at
${\rho}^{FNM}_{RSB}\approx 0.380(5)$. 
Fitting these curves 
leads unambiguously to a non-zero large $L$
value for the order parameter when ${\rho}=0.37$ 
and to a zero value when ${\rho}=0.39$.
In the SNM (see inset of Fig.~\ref{fig_rho_ex}), we find 
${\rho}^{SNM}_{RSB}\approx 0.235(5)$ and 
the extrapolations 
away from this point behave as in the FNM. 
Note that, within statistical errors, the curves have a common
crossing point. Furthermore, 
just as for $\rho =0$, up to the transition point
the fraction of sponge-like events {\it increases} with $L$.
It thus seems most likely
that these excitations survive as 
$L \to \infty$ if $\rho < \rho_{RSB}$. This is
as expected in the mean field approach, but not
in the droplet/scaling pictures.
However, the transition point 
${\rho}_{\rm{RSB}}$ is close if not equal to
${\rho}_{\rm{F}}$ in both of our models, so it seems to us that
there is no mixed phase having both sponge excitations 
and positive magnetization.
This would be in contrast to the mean field prediction.

\paragraph*{Domain wall wandering and ${\theta}_{\rm{DW}}$ ---}
We now study directly the spin glass ordering, be-it
with or without RSB, so we no longer resort to $O(1)$ energy sponges.
It is well known that spin glass ordering is sensitive
to perturbations; the standard probe for this
consists in comparing periodic and anti-periodic boundary 
conditions~\cite{BrayMoore86},
thereby creating an {\it interface} 
in the system. The typical
energy change for such a modification of boundary conditions
should grow more slowly than $L^2$ in the spin glass phase,
with $O(1)$ energy changes arising not too rarely,
for instance with a probability going as an inverse power of $L$.
Another important aspect of 
such an interface
is associated with its position space properties:
in a spin glass phase, it is ``space spanning''. This 
can be made precise algorithmically by considering whether
the interface winds around the whole lattice. We take this
behavior to be an indication of the ``fragility''
of the ground state; we thus consider that we are in
a spin glass phase when there is a positive probability
that the interface is ``spongy'', {\it i.e.}, winds around the 
lattice.

\begin{figure}
\centerline{\hbox{\epsfig{figure=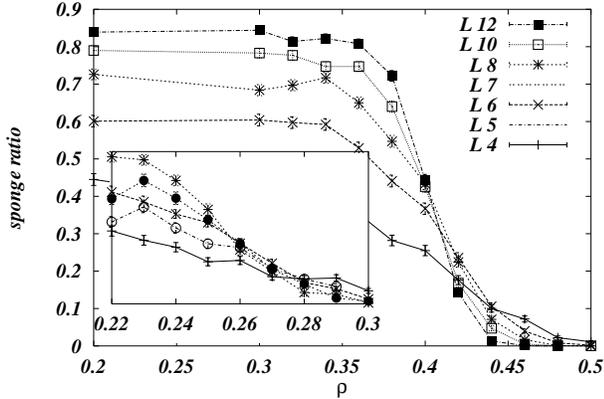,width=8cm}}}
\caption{Fraction of spongy interfaces versus $\rho$ in the FNM. (The inset
shows the data for the SNM.) The crossing points drift to 
the left, suggesting ${\rho}_{\rm{DW}} \approx {\rho}_{\rm{RSB}}$.}
\label{fig_rho_dw}
\end{figure}

We begin with the FNM.
In Fig.~\ref{fig_rho_dw} we show the fraction of instances
where the interface is spongy.
The data suggest a transition at ${\rho}_{\rm{DW}}{\leq}0.4$. However 
there is a systematic drift of the crossing points to
the left so the most straight-forward 
interpretation of our results 
is ${\rho}_{\rm{RSB}}={\rho}_F={\rho}_{\rm{DW}}$;
then the spin glass ordering disappears when the ferromagnetic 
ordering appears, the two orderings do not co-exist and there is 
{\it no} mixed phase. 

If on the other hand one really believes mean field to be a 
good guide, one can assume that the mixed phase is present 
but only in a very small window, something like 
${\rho}\in [0.38,0.40]$. To make 
such a window larger and thus to give creadence
in favor of a mixed phase, we
increase the connectivity of the lattice
from $6$ to $18$ by going to the SNM. There, our analysis
leads to ${\rho}_{\rm{DW}} \leq 0.255$; thus in this
``improved'' model, the most optimistic estimate for 
the window would be ${\rho}\in [0.236,0.255]$. However 
just as in the FNM, the crossing points of the curves drift 
and again ${\rho}_{\rm{DW}} - {\rho}_{\rm{RSB}}$ is very small if
not zero, a result quite puzzling from the mean field perspective,
not to mention the fact that RSB seems to be excluded in this window. 

To pursue the search for a mixed phase, we have studied the
FNM in $d=4$ for sizes $L \le 5$.
The behavior is similar to the case $d=3$, and in particular we obtain
${\rho}_F=0.315(5)$ and ${\rho}_{\rm{RSB}}=0.325(5)$
(data not shown). But the crossing
points of the Binder cumulant which determine
${\rho}_F$ now shift towards larger values
as $L$ increases. As a result, there is
no compelling evidence for a mixed phase and
probably $\rho_{DW}=\rho_{F}=\rho_{RSB}$. 

\paragraph*{Discussion and conclusions ---} In spite of our attempts
to render finite dimensional models mean-field-like,
we are unable to confirm the mean field prediction
that RSB and ferromagnetic order can co-exist.
Since we also find quite clear signals for RSB 
via the existence of spongy clusters having $O(1)$ energies
in the whole unmagnetized region, we cannot explain our results
using the droplet/scaling theories either.
(There is no RSB in those approaches.) Interestingly, all
the results we obtain
are compatible with the ``Trivial-Non-Trivial'' or TNT 
picture~\cite{KrzakalaMartin00,PalassiniYoung00a}. 
In that picture, there are
system-size excitations whose energies are $O(1)$; this leads to RSB,
{\it i.e.}
a non-trivial distribution of spin overlaps.
Furthermore, they are also compact,
with a surface fractal dimension $d_s<d$, leading to
a trivial distribution of link overlaps.
This TNT picture forbids a mixed phase as follows.
Assume that the local magnetization density (in a big enough
sub-volume of the whole) is self-averaging and 
equal to the global magnetization density. Activating
compact system-size excitations will not affect the
local magnetization but will change the global one unless
$m=0$. A well behaved magnetization with TNT then forces
$m=0$, {\it i.e.}, no mixed phase.

Other numerical studies, in particular based on Monte-Carlo 
simulations, are needed. On the analytical side,
a computation of the phase diagram on the Bethe lattice 
with RSB~\cite{Mezardparisi01}, or field theoric approaches for
the mixed phase in the spirit of~\cite{PimentelTemesvari02}, would 
be welcome. Finally, experimentally, it may be possible
to test whether magnetization and spin glass ordering
can co-exist {\it in equilibrium}; however this will require careful
checks that equilibrium is indeed reached.

\paragraph*{Acknowledgements ---} We thank M. M\'ezard, E. Vincent,
O. White and A.P. Young for very useful discussions. 
FK acknowledges support from the MENRT. The 
LPTMS is an Unit\'e de Recherche de 
l'Universit\'e Paris~XI associ\'ee au CNRS.
\bibliographystyle{prsty}
\bibliography{../../../Bib/references}

\end{document}